\documentclass[prl,twocolumn,showpacs,amsmath,superscriptaddress,psfig,amssymb]{revtex4}
\usepackage{graphicx}% Include figure files
\usepackage{dcolumn}% Align table columns on decimal point
\usepackage{bm}% bold math

\begin{document}

%\preprint{APS/123-QED}

\title{Non-Fermi liquid states in the pressurized CeCu$_2$(Si$_{1-x}$Ge$_x$)$_2$ system:\\two critical points}

\author{H. Q. Yuan}
 \altaffiliation [Present address: ]
{Department of Physics, University of Illinois at Urbana and
Champaign, 1110 West Green Street, Urbana, IL 61801}
\email{yuan@mrl.uiuc.edu.}\affiliation {Max-Planck-Institute for
Chemical Physics of Solids, N\"{o}thnitzer Stra\ss e 40, 01187
Dresden, Germany}

\author{F. M. Grosche}
\affiliation {Department of Physics, Royal Holloway, University of
London, Egham TW20 0EX, UK}

\author{M. Deppe}
\author{G. Sparn}
\author{C. Geibel}
\author{F. Steglich}
\affiliation {Max-Planck-Institute for Chemical Physics of Solids,
N\"{o}thnitzer Stra\ss e 40, 01187 Dresden, Germany}

\date{\today}% It is always \today, today,

\begin{abstract}
\noindent In the archetypal strongly correlated electron
superconductor CeCu$_2$Si$_2$ and its Ge-substituted alloys
CeCu$_2$(Si$_{1-x}$Ge$_{x}$)$_2$ two quantum phase transitions --
one magnetic and one of so far unknown origin -- can be crossed as
a function of pressure \cite{Yuan 2003a}.  We examine the
associated anomalous normal state by detailed measurements of the
low temperature resistivity ($\rho$) power law exponent $\alpha$.
At the lower critical point (at $p_{c1}$, $1\leq\alpha\leq 1.5$)
$\alpha$ depends strongly on Ge concentration $x$ and thereby on
disorder level, consistent with a Hlubina-Rice-Rosch scenario of
critical scattering off antiferromagnetic fluctuations. By
contrast, $\alpha$ is independent of $x$ at the upper quantum
phase transition (at $p_{c2}$, $\alpha\simeq 1$), suggesting
critical scattering from local or $Q=0$ modes, in agreement with a
density/valence fluctuation approach.

\end{abstract}

\maketitle

%%%%%%%%%%%%%%%%%%%%%%%%%%%%%%%%%%%%%%%%%%%%%%%%%%%%%%%%%%%%%%%%%%%%%%
%%                                                                  %%
%%      Introduction                                                %%
%%                                                                  %%
%%%%%%%%%%%%%%%%%%%%%%%%%%%%%%%%%%%%%%%%%%%%%%%%%%%%%%%%%%%%%%%%%%%%%%

\noindent Amongst the cerium based f-electron compounds, the
superconductor CeCu$_2$Si$_2$ \cite{Steglich 1979} takes a special
place. The difficulty in growing high quality samples with
reproducible properties and the diversity of observed low
temperatures states have long complicated and delayed a
theoretical description of this intriguing material. After more
than 25 years of intensive study, its key properties are gradually
being understood. Initial confusion about the ground state
properties of CeCu$_2$Si$_2$ samples -- some magnetic, some
superconducting -- can now be attributed unambiguously to the
delicate positioning of this material close to a magnetic quantum critical point (QCP)
\cite{Gegenwart 1998}. The precise nature of the incipient
magnetism in ambient-pressure CeCu$_2$Si$_2$ has recently been
determined as incommensurate spin density wave order
\cite{Stockert04}. Superconductivity in low pressure
CeCu$_2$Si$_2$ now appears amenable to an analysis along the same
lines as in other Ce-based heavy fermion (HF) compounds on the threshold of
magnetism \cite{Mathur 1998}, in terms of magnetically mediated
pairing.  The evolution of CeCu$_2$Si$_2$ under high pressure,
however, has opened up new questions.

The pressure dependence of the superconducting transition
temperature $T_c$ in CeCu$_2$Si$_2$ \cite{Thomas 1993, Bellarbi
1984} and in its isoelectronic sister compound CeCu$_2$Ge$_2$
\cite{Jaccard 1999} is very different from that observed in other
Ce-based HF compounds, such as CePd$_2$Si$_2$ and CeIn$_3$. In
CeCu$_2$Si$_2$, $T_c$ is nearly pressure independent up to about
$2 ~{\rm GPa}$ away from the antiferromagnetic (AFM) QCP (at $p_{c1}$) and then
increases to a maximum value about 3-4 times that at $p_{c1}$.

\begin{figure}
\centering
\includegraphics[width=0.95\columnwidth]{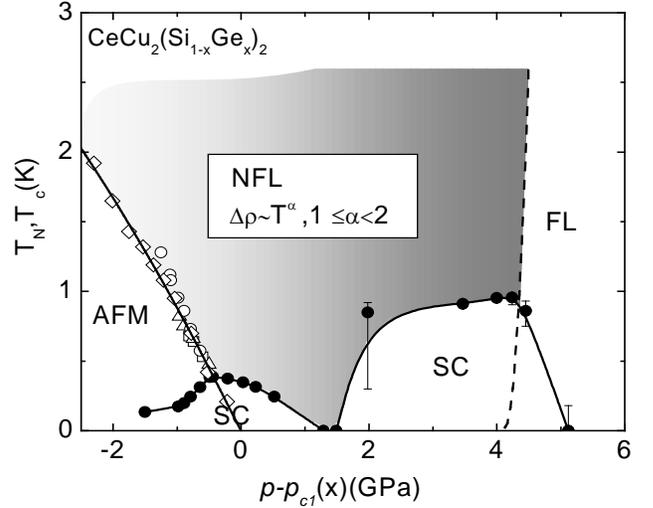}
\caption{The combined $p-T$ phase diagram for
CeCu$_2$(Si$_{1-x}$Ge$_x$)$_2$ ($T_N$: $x=$0.25 ($\diamond$), 0.1
($\circ$), 0.05 ($\bigtriangleup$), 0.01 ($\square$); $T_c$: $x=$0.1
($\bullet$)). }
\label{PhaseDia}
\end{figure}

To understand the origin of this phase diagram, we have recently
performed a study on a series of partially Ge-substituted single
crystals CeCu$_2$(Si$_{1-x}$Ge$_x$)$_2$. Due to the weakening of
superconductivity by the increased impurity scattering associated with
Ge substitution \cite{Yuan 2004} (which widens the lattice and is
counterbalanced by applying hydrostatic pressure), the broad and
continuous superconducting range previously observed in the $p-T$
phase diagram of pure CeCu$_2$Si$_2$ and CeCu$_2$Ge$_2$ breaks up into
two disconnected superconducting domes \cite{Yuan 2003a}. The
low-pressure superconducting dome occurs around an AFM QCP, suggesting
magnetically mediated pairing, while the high-pressure superconducting
dome straddles a weak first-order volume collapse
(Fig.~\ref{PhaseDia}) indicative of a second quantum phase transition
(QPT) at high pressure. In this letter, we elucidate the nature of the
two QPTs by studying their anomalous normal-state behavior.

%%%%%%%%%%%%%%%%%%%%%%%%%%%%%%%%%%%%%%%%%%%%%%%%%%%%%%%%%%%%%%%%%%%%%%
%%                                                                  %%
%%      Experimental Methods                                        %%
%%                                                                  %%
%%%%%%%%%%%%%%%%%%%%%%%%%%%%%%%%%%%%%%%%%%%%%%%%%%%%%%%%%%%%%%%%%%%%%%

\begin{figure}
\centering
\includegraphics[width=0.95\columnwidth]{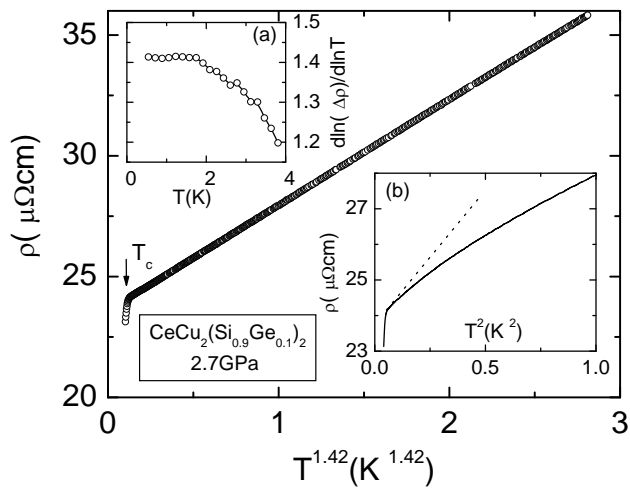}
\caption{The fit of the resistivity by $\rho=\rho_0+AT^\alpha$ for
CeCu$_2$(Si$_{0.9}$Ge$_{0.1}$)$_2$ at $p=2.7$ GPa. }
\label{Normal27}
\end{figure}

Single crystals of CeCu$_2$(Si$_{1-x}$Ge$_x$)$_2$ have been
prepared by a flux growth method in excess Cu. High sensitivity,
AC four-point measurements of the electrical resistivity were
carried out in Bridgman anvil ($p<10 ~{\rm GPa}$) and
piston-cylinder ($p<3.5 ~{\rm GPa}$) devices down to 200 mK in an
adiabatic demagnetization cooler and down to 50 mK in an Oxford
Instruments dilution refrigerator. The normal state behavior of
our samples has been analyzed by fitting the low temperature
normal state resistivity as $\rho=\rho_0+AT^\alpha$ up to an
adjustable maximum temperature $T_{max}$. The resulting residual
resistivity $\rho_0$ can be used to extract the temperature
dependence of $\alpha$ by taking the logarithmic derivative
$\alpha(T)=d\ln(\rho(T)-\rho_0)/d\ln T$, as illustrated in
Fig.~\ref{Normal27} \cite{Yuan 2003b}. Both methods are iterated
until convergence in $\alpha$ and $T_{max}$ is achieved. We note
that $T_{max}$ -- which represents the range of validity of the
asymptotic low-$T$ power law behavior -- depends on Ge
concentration and on external pressure. It increases from about 2
K at low $p$ to 10 K at $p_{c2}$ (indicated by darkness of shading
in Fig.~\ref{PhaseDia}).

%%%%%%%%%%%%%%%%%%%%%%%%%%%%%%%%%%%%%%%%%%%%%%%%%%%%%%%%%%%%%%%%%%%%%%
%%                                                                  %%
%%      To begin with: an overview over the whole landscape         %%
%%                                                                  %%
%%%%%%%%%%%%%%%%%%%%%%%%%%%%%%%%%%%%%%%%%%%%%%%%%%%%%%%%%%%%%%%%%%%%%%

Fig.~\ref{PhaseDia} summarizes our present knowledge of the ordered
phases of the CeCu$_2$(Si$_{1-x}$Ge$_x$)$_2$ system. It has been
constructed by shifting the pressure scale for each Ge concentration
by the respective lower critical pressure, $p_{c1}$, at which the AFM
transition temperature $T_N$ extrapolates to zero. The critical
pressure $p_{c1}$ is about 1.4, 1.5, 1.5 and 2.4 GPa for $x$=0.01,
0.05, 0.1 and 0.25, respectively. Due to Cu/Si site exchange and
possible sample inhomogeneities, the value of $p_{c1}$ becomes less
regular for small $x$. Following such a pressure shift, the abscissa
can be approximately regarded as a volume scale \cite{Yuan 2003a}.
This observation is consistent with the existence of an AFM QCP in
CeCu$_2$(Si$_{1-x}$Ge$_2$)$_2$, and indicates that the magnetic QCP
exists at a unique volume of the unit cell. At very high pressures, as
the system is tuned out of the HF state and into an intermediate
valence state, it undergoes an isostructural first-order volume
collapse, possibly analogous to the $\gamma-\alpha$ transition in
elemental Ce. The likely pressure dependence of this transition is
schematically indicated by a dashed line in
Fig.~\ref{PhaseDia}. Indeed, a weak first-order volume-collapse line
with an apparently low-lying critical end point have been observed
around a second QPT in CeCu$_2$Ge$_2$, where $T_c$ reaches a maximum
value \cite{Onodera 2002}.

\begin{figure}
\centering
\includegraphics[width=0.8\columnwidth]{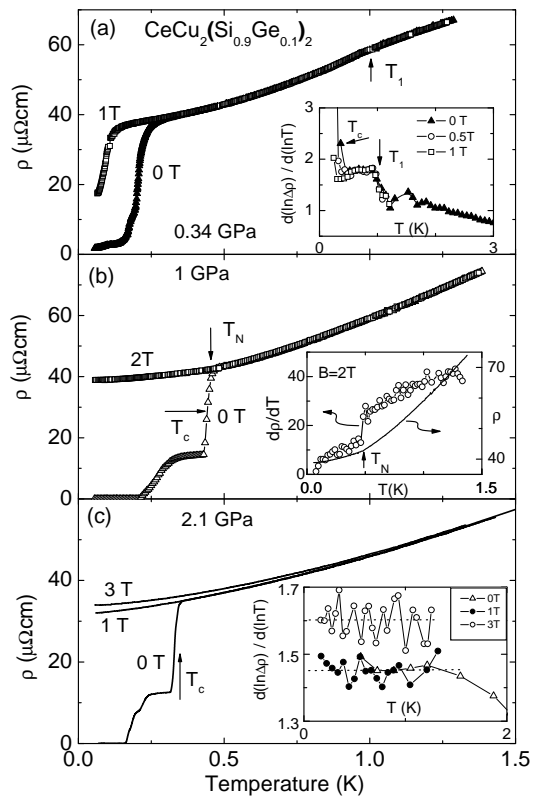}
\caption{The electrical resistivity $\rho(T)$ for
CeCu$_2$(Si$_{0.9}$Ge$_{0.1}$)$_2$ at various magnetic fields and
hydrostatic pressures, demonstrating three types of ground states
(see text). } \label{AtPc1}
\end{figure}

The pressure dependence of the N\'{e}el temperature $T_N$ and the
volume collapse transition divide the phase diagram into three
regions: the antiferromagnetically ordered state below $p_{c1}$,
the intermediate valence range above $p_{c2}$, and the more
complex region in between the two QPTs.

%%%%%%%%%%%%%%%%%%%%%%%%%%%%%%%%%%%%%%%%%%%%%%%%%%%%%%%%%%%%%%%%%%%%%%
%%                                                                  %%
%%      First result: around p_c1                                   %%
%%                                                                  %%
%%%%%%%%%%%%%%%%%%%%%%%%%%%%%%%%%%%%%%%%%%%%%%%%%%%%%%%%%%%%%%%%%%%%%%

Focussing initially on the normal state around the low pressure
AFM QCP, we note that different low temperature states can be
obtained in ambient pressure CeCu$_2$Si$_2$ by deliberately
choosing the composition of the melt to be slightly off
stoichiometry or by suitable heat treatments \cite{Steglich 1995,
Lang 1999}. On the other hand, very similar ground states can be
achieved in slightly Ge-substituted samples by applying
hydrostatic pressure. This allows us to study the magnetic QCP in
greater detail. As an example, Fig.~\ref{AtPc1} shows three
possible cases: (a) {\em Magnetic} ($T_N > T_c$). At $p=0.34$ GPa,
CeCu$_2$(Si$_{0.9}$Ge$_{0.1}$)$_2$ experiences a magnetic
reorientation transition at $T_1\simeq 1$ K (the initial AFM
transition is at $T_N\simeq 1.4$ K), followed by a superconducting
transition at $T_c\simeq 0.2$ K. Upon applying a magnetic field,
superconductivity is quickly suppressed, but the magnetism is much
more robust (inset of Fig.~\ref{AtPc1}a). The exponent $\alpha$
decreases with increasing temperature and above $T_N$, $\alpha$
remains $<2$ (inset of Fig.~\ref{AtPc1}a). The non-$T^2$ form of
$\rho(T)$ above $T_N$ agrees with thermodynamic properties
\cite{Gegenwart 1998}, pointing at a non-Fermi liquid (NFL) normal
state above $T_N$. (b) {\em Superconducting/Magnetic} ($T_N\leq
T_c$). In this case, the magnetic transition is masked by
superconductivity, but reappears as superconductivity is
suppressed below $T_N$ by a magnetic field. (inset of
Fig.~\ref{AtPc1}b). (c) {\em Superconducting}. No magnetic
transition can be observed even when superconductivity is
suppressed by a magnetic field. Generally, the magnetic field has
little effect on the normal state as long as the field is below
the upper critical field $B_{c2}$ (insets of Figs.~\ref{AtPc1}a
and \ref{AtPc1}c). When the magnetic field exceeds $B_{c2}$, the
exponent $\alpha$ gradually increases with increasing magnetic
field (Fig.~\ref{AtPc1}c).

%%%%%%%%%%%%%%%%%%%%%%%%%%%%%%%%%%%%%%%%%%%%%%%%%%%%%%%%%%%%%%%%%%%%%%
%%                                                                  %%
%%      Second result: p-dependence of exponent                     %%
%%                                                                  %%
%%%%%%%%%%%%%%%%%%%%%%%%%%%%%%%%%%%%%%%%%%%%%%%%%%%%%%%%%%%%%%%%%%%%%%

At comparatively low pressures $p\simeq p_{c1}$, the interplay between
superconductivity and magnetism in the CeCu$_2$(Si/Ge)$_2$ system
exhibits a similar structure to what is seen in other quantum
critical Ce based HF superconductors, such as CePd$_2$Si$_2$.
However, the question arises how the normal state develops with
increasing distance from the AFM QCP, and how it connects up with
the volume collapse QPT at high pressure.

\begin{figure}
\centering
\includegraphics[width=0.8\columnwidth]{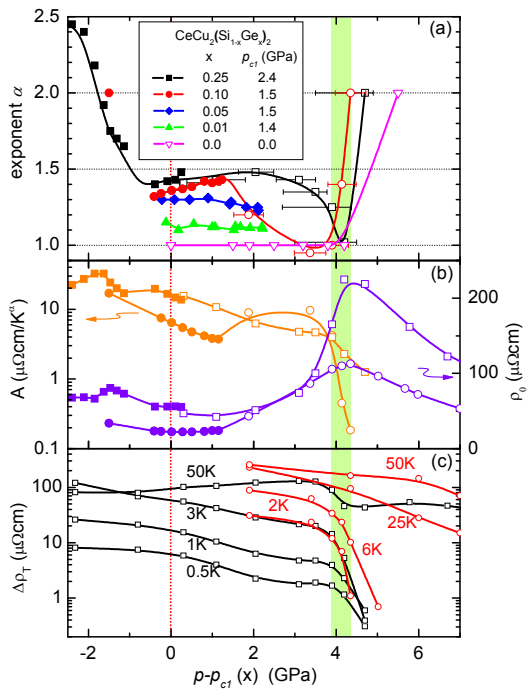}
\caption{(Color online). The pressure dependence of (a) the
resistivity exponent $\alpha$ ($x=0$ is from Ref. \cite{Bellarbi
1984}); (b) the resistivity $A$ coefficient and the residual
resistivity $\rho_0$; and (c) the resistivity isotherms
$\Delta\rho_T(p)$ ($=\rho(p,T)-\rho_0(p)$) at various temperatures
for CeCu$_2$(Si$_{1-x}$Ge$_x$)$_2$ ($x=0.25$ ($\square$), 0.1
($\circ$), 0.05 ($\diamond$), 0.01 ($\vartriangle$) and 0.0
($\triangledown$)). The filled symbols represent the samples
measured in clamped pressure cells and the empty ones are from
Bridgman anvil cells.} \label{Exponents}
\end{figure}

Examining the evolution of the resistivity exponent $\alpha$
across the $p-T$ phase diagram (Fig. 4a), we note the following
key points: (i) At the AFM QCP (at $p_{c1}$), the exponent
$\alpha$ reaches a local minimum. The value of $\alpha$ at
$p_{c1}$ ranges between 1 and 1.5 and increases with increasing
Ge-content $x$. (ii) The exponent $\alpha$ reaches a second
minimum in the high-pressure superconducting regime, approaching
$\alpha\approx 1$ around the volume collapse transition at
$p_{c2}$ ($\Delta p\sim 4$ GPa). Maximum $T_c$ is accompanied in
CeCu$_2$Si$_2$ and its Ge-substituted alloys by an extended
$T$-linear form of the resistivity -- {\em independent of Ge
content (and of the associated disorder)}. Upon further increasing
pressure above $p_{c2}$, Fermi-liquid behavior ($\alpha=2$) is
rapidly recovered. (iii) In between the two QPTs, for
$p_{c1}<p<p_{c2}$, NFL behavior with $1\leq \alpha <2$ survives
over a broad range in pressure (about 4 GPa). For small Ge
concentrations (e.g. $x$ =0, 0.01 and 0.05), $\alpha$ is nearly
pressure independent above $p_{c1}$. However, $\alpha$ goes
through a local maximum at intermediate pressure for larger $x$
($x = 0.1$ and 0.25).

As in other quantum critical HF compounds, current theories can
only account qualitatively for the anomalous normal state observed
in CeCu$_2$(Si$_{1-x}$Ge$_x$)$_2$. At the AFM QCP,
spin-fluctuation theories \cite{ Hertz 1976, Millis 1993, Moriya
1995, Lonzarich 1997} predict $\alpha=1.5$ and $\alpha=1$ for 3D-
and 2D- spin fluctuations, respectively, while our measured
exponents are sample dependent and lie between these two extremes.
The observed increase of $\alpha$ with increasing disorder ($1\leq
\alpha\leq 1.5$) may, however, be explained within a generalised
Hlubina-Rice type hot-spot/cold-spot scenario, e.g. \cite{Rosch
1999}. Such an approach takes into account both the
short-circuiting of critical scattering at large wavevector ${\bf
q=Q}$ (connecting ``hot'' regions of the Fermi surface) by
``cold'' regions, and the influence of impurity scattering, which
is present at all ${\bf q}$.

%%%%%%%%%%%%%%%%%%%%%%%%%%%%%%%%%%%%%%%%%%%%%%%%%%%%%%%%%%%%%%%%%%%%%%
%%                                                                  %%
%%      Third result: High pressure normal state                    %%
%%                                                                  %%
%%%%%%%%%%%%%%%%%%%%%%%%%%%%%%%%%%%%%%%%%%%%%%%%%%%%%%%%%%%%%%%%%%%%%%

The presence of a second QPT at $p_{c2}$ holds the key for
understanding the unusual pressure dependence of the resistivity
exponent in between $p_{c1}$ and $p_{c2}$. In Figs.~4b and 4c, the
pressure dependence of the $A$ coefficient in
$\Delta\rho=AT^\alpha$ and the resistivity isotherms $\Delta\rho_T
(p)$ ($=\rho(p,T)-\rho_0(p)$) at various temperatures are shown
for the samples with $x=0.1$ and $x=0.25$. The collapse of $\Delta
\rho_T(p)$ (at $T<10 ~\rm K$) and of $A(p)$ on crossing the upper
critical pressure $\Delta p=p_{c2}-p_{c1}(x)\simeq 4$ GPa,
indicates a transition from the HF state to an intermediate
valence state at $p_{c2}$. This valence transition may be
accompanied by an isostructural, weak first-order volume collapse,
as suggested by x-ray diffraction experiments on CeCu$_2$Ge$_2$
\cite{Onodera 2002}. At temperatures exceeding 10 K, the drop in
the resistivity isotherms at $p_{c2}$ weakens
(Fig.~\ref{Exponents}c), and it vanishes below 50 K. These data
suggest that the first order transition line associated with the
putative density/valence change at $p_{c2}$ reaches its critical
end point at a very low temperature, less than 50 K, explaining
also why various past attempts to observe the volume collapse in
CeCu$_2$(Si/Ge)$_2$ by high pressure x-ray diffraction at room
temperature have remained unsuccessful.

A weak volume collapse transition at $p_{c2}$ is expected to be
accompanied by large amplitude fluctuations of the lattice density and
consequently of the local charge distribution (i.e. the
valence). Charge carrier scattering is modified in the presence of
these fluctuations, giving rise to an anomalous temperature dependence
of $\rho(T)$, provided that the fluctuation relaxation rate reaches
down to low enough energies. In the most detailed scenario so far,
proposed by Miyake \cite{Holmes 2004}, non-dispersive (local), but
nearly critical valence fluctuations are invoked to explain the linear
$T$-dependence of $\rho(T)$ at $p_{c2}$, essentially as a consequence
of the equipartition theorem. It is as yet unclear whether this
approach can also explain the absence of a giant heat capacity or
$A$-coefficient peak, which would be expected in the presence
of very low-lying excitations spread over large portions of the
Brillouin zone, as well as the occurrence of superconductivity, which
usually requires a non-local pair-forming interaction. Density or
valence fluctuations peaked at ${\bf q}=0$, whether dispersive or nearly local, would
however offer an explanation for the observed disorder-level
independent power-law exponent at $p_{c2}$, because in this case the
entire Fermi surface can be considered ``hot''. In contrast to the AFM
QCP at $p_{c1}$, where a hot-spot/cold-spot scenario accounted at
least qualitatively for the impurity-level dependence of $\alpha$, the
$T$-linear resistivity obtained from a density or valence-fluctuation
model should then be robust against the level of disorder -- in
agreement with our experimental findings.

Second to superconductivity, arguably the most dramatic phenomenon
in the CeCu$_2$(Si/Ge)$_2$ system is the enormous enhancement of
the residual resistivity $\rho_0$ around $p_{c2}$ (Fig. 4b), which
contrasts starkly with the weak {\em minimum} in $\rho_0$ at
$p_{c1}$. The origin of this distinct peak in $\rho_0(p)$ has been
proposed to lie in a strongly pressure dependent impurity
scattering cross-section, as $p_{c2}$ is approached. Here, the
problem lies in the computed logarithmic dependence of $\rho_0$ on
distance from the critical point \cite{Miyake 2002}, coupled with
the first order nature of the volume collapse transition at low
$T$.  An alternative approach to the state of CeCu$_2$(Si/Ge)$_2$
near $p_{c2}$ may consider the likely phase separation into
low-density (HF) and high-density (intermediate valent) domains,
populated by heavy and light carriers, respectively, in distant
analogy with the mechanism underlying Giant Magnetoresistance. On
the assumption that light quasi-particles cannot propagate in
heavy-fermion domains and conversely, heavy quasi-particles
scatter strongly in the intermediate-valent (high density)
domains, CeCu$_2$Si$_2$ is expected to turn opaque to electrical
transport over a narrow region surrounding $p_{c2}$, leading to
the observed pronounced maximum in $\rho_0(p)$.

In contrast to stoichiometric CeCu$_2$Si$_2$, in which a
quasi-linear $T$-dependence of the resistivity extends over the
entire region between $p_{c1}$ and $p_{c2}$, the resistivity
exponent $\alpha$ in Ge-substituted CeCu$_2$Si$_2$ single crystals
reaches two distinct minima at $p_{c1}$ and $p_{c2}$.  These
results indicate that the apparent critical {\em region} in the
$p-T$ phase diagram of stoichiometric CeCu$_2$Si$_2$ is a result
of two critical {\em points}, each surrounded by a pressure range
in which $\alpha$ is low. We arrive, then, at a picture analogous
to the explanation for the wide superconducting range in
stoichiometric CeCu$_2$Si$_2$, which is attributed to the merger
of the two superconducting domes in Ge-substituted
CeCu$_2$(Si/Ge)$_2$: the interplay of two QPTs results in the
unusual pressure dependence of both superconductivity and normal
state behavior in CeCu$_2$Si$_2$. While the AFM critical point at
$p_{c1}$ is similar in nature to that in other Ce based HF
compounds, the precise nature and origin of the QPT at $p_{c2}$ is
still unclear. Some of its consequences -- the colossal pressure
dependence of $\rho_0$ and the linear, disorder-level independent
$T$-dependence of $\rho$ -- are, however, clearly established and
invite further theoretical investigation.

We thank P. Gegenwart, G. G. Lonzarich, K. Miyake, P. Monthoux, J.
A. Mydosh, and M. B. Salamon for useful discussions. HQY also
acknowledges the ICAM postdoctoral fellowship.

\end{document}